# *Nanocrystalline silicon optomechanical cavities*

D. Navarro-Urrios[1,2], N.E. Capuj[3,4], J. Maire[1], M. Colombano[1,5], J. Jaramillo-Fernandez[1], E. Chavez-Angel[1], L. L. Martin[6], L. Mercadé[6], A. Griol[6], A. Martínez[6], C. M. Sotomayor-Torres[1,7], J. Ahopelto[8]

*[1]Catalan Institute of Nanoscience and Nanotechnology (ICN2), CSIC and The Barcelona*

*Institute of Science and Technology, Campus UAB, Bellaterra, 08193 Barcelona,*

*Spain*

*[2]MIND-IN2UB, Departament d'Electrònica, Facultat de Física, Universitat de Barcelona, Martí i Franquès 1, 08028 Barcelona, Spain*

*[3]Depto. Física, Universidad de la Laguna, La Laguna, Spain*

*[4]Instituto Universitario de Materiales y Nanotecnología, Universidad de La Laguna, 38200 San Cristóbal de La Laguna, Spain.*

*[5]Depto. Física, Universidad Autónoma de Barcelona, Bellaterra, 08193 Barcelona, Spain.*

*[6]Nanophotonics Technology Center, Universitat Politècnica de Valencia, Spain*

*[7]Catalan Institute for Research and Advances Studies ICREA, Barcelona, Spain*

*[8]VTT Technical Research Centre of Finland Ltd, P.O. Box 1000, FI-02044 VTT, Espoo, Finland*

Abstract:

Silicon on insulator photonics has offered a versatile platform for the recent development of integrated optomechanical circuits. However, there are some constraints such as the high cost of the wafers and limitation to a single physical device level. In the present work we investigate nanocrystalline silicon as an alternative material for optomechanical devices. In particular, we demonstrate that optomechanical crystal cavities fabricated of nanocrystalline silicon have optical and mechanical properties enabling non-linear dynamical behaviour and effects such as thermo-optic/free-carrier-dispersion self-pulsing, phonon lasing and chaos, all at low input laser power and with typical frequencies as high as 0.3 GHz.

Introduction

Interaction between electromagnetic and mechanical waves can be significantly enhanced when the waves are confined into high-quality factor optical cavities. This emerging field, known as cavity optomechanics, has become a powerful framework to observe a plethora of new phenomena both in the classical and quantum domains.[1,2] Non-linear optomechanics is gaining interest because of the novel features, including phonon lasing,[3] chaos[4] and quadratic readout of displacement.[5] Optomechanical cavities (OMC) to boost interaction between light and sound can be achieved in many ways, from ultra-high quality factor (Q) silica micro-toroids[6,7] or spheres[8] to high-finesse Fabry-Perot cavities.[9] Recently, cavities created in planar semiconductor films by tailoring periodic patterns either in one or two-dimensions have become particularly popular.[10-12] The patterning is made by well-established top-down fabrication tools, typically electron-beam lithography and dry etching, which have several advantages over alternative approaches, such as design flexibility, scalability at different frequency regimes or dimensions, and integration with electronics and coupling to optical fibres.

Since light and sound propagation velocities in solid materials differ by about five orders of magnitude, OMCs support typically GHz-range mechanical resonances for phonons, while localized photons are in the near-infrared regime, ~200 THz.[13] Using high-Q cavities, strong interaction between optical and mechanical waves in an OMC has been demonstrated in many crystalline and polycrystalline materials, such as silicon nitride ($Si_3N_4$),[14] gallium arsenide (GaAs),[15] aluminium nitride (AlN),[16] diamond,[17] and, notably, in crystalline silicon (Si).[10,18]

Silicon is particularly interesting as the core material for on-chip cavity optomechanics. Silicon on insulator (SOI) substrates are massively used in photonic integrated circuits due to the high index of refraction and negligible losses at the telecom wavelengths. In addition, the existing technology for sub-micron patterning using processes compatible with CMOS technology allows large-scale production. Achieving negligible optical losses typically requires single crystalline silicon (c-Si), and researchers have reported remarkable results for ultra-low loss silicon waveguides[19,20] or ultra-high Q cavities.[21] The same applies to cavity optomechanics: Large optical and mechanical Q factors have been observed in cavities created in c-Si OMCs.[10,11,18] In general, crystallinity of the core material is highly recommended in linear applications when the propagation or confinement losses are to be minimized. However, the situation is different in non-linear applications, where an extra amount of losses can be compensated by non-linear properties of a non-crystalline material. In the case of silicon, non-linear thermal and free-carrier effects, which can be used to dynamically tune the optical properties of Si, are different in crystalline, polycrystalline and amorphous layers [22-24]. Indeed, amorphous and polycrystalline Si provide shorter recombination lifetimes of free carriers than c-Si, resulting in larger and faster non-linear effects.[25,26] This has been used to implement on-chip amorphous and polycrystalline Si-based all-optical devices operating much faster (> 10 GHz) than their c-Si counterparts.[23,24,27,28] Nanocrystalline silicon (nc-Si) is a specific type of polycrystalline Si that is widely used in MEMS production due to the relatively easy tuning of the grain-size and stress, i.e., mechanical and optical properties, and electrical properties.[29] Here we assess for the first time the applicability of nc-Si as a material for OMCs with the special focus on non-linear optomechanical features.

Results

In this work we study experimentally the optomechanical properties of a 1D OMC made of a suspended nc-Si film. In particular, we use the same nominal OMC design (see Fig. 1(a)) that was used earlier to demonstrate a phononic bandgap,[18] phonon lasing[3] and complex non-linear dynamics such as chaos[4] in OMCs made of c-Si. The 1D cavity is constructed of square unit cells containing a hole in the middle and two symmetric stubs on the sides. The defect region of the OMCs consists of 12 central cells in which the pitch (a), the radius of the hole (r) and the stubs length (d) are decreased in a quadratic way towards the centre. The maximum reduction of the parameters is denoted by $\Lambda$. A 10 period mirror is included on both sides of the defect region. The nominal geometrical values of the cells of the mirror are a=500 nm, r=150 nm and d=250 nm. The total number of cells is 32 and the whole device length is ~15 µm. All the results presented in this work correspond to the structure with $\Lambda$=0.8.

The nominal structure of the nc-Si SOI-like wafers was designed to have a 220 nm thick nc-Si film on a 1000 nm thick oxide layer. The fabrication process included growth of a thick $SiO_2$ by wet oxidation at 1050 °C and a layer of amorphous Si at 574 °C by chemical vapour deposition (CVD). Amorphous Si deposited by CVD is typically under compressive stress and is not as such suitable for released structures. To convert the compressive stress to tensile, the wafers were annealed at 950 °C, which resulted in a few tens of MPa tensile stress. The annealing step transforms the amorphous film to nanocrystalline with the grain size ranging from a few nm to 100-200 nm.[20] The measured layer thicknesses after annealing were 1013 nm and 223 nm for the $SiO_2$ film and nc-Si film, respectively. Fig. 1(b) shows Raman spectra, taken using a 532 nm pump laser, of the fabricated layer stack together with that of a reference crystalline silicon sample. The typical optical-phonon mode of crystalline Si appears in both cases at 520 cm$^{-1}$. The fabricated layer shows a weak asymmetric broadening at smaller energies, which can be associated to the presence of a minor amorphous phase, grain boundaries and nanocristallites of different sizes.[30]

Suspended beams with OMCs were processed on the nc-Si wafers in the same way as in the case of the cavities on SOI wafers (fabrication details are described in Ref. 18).

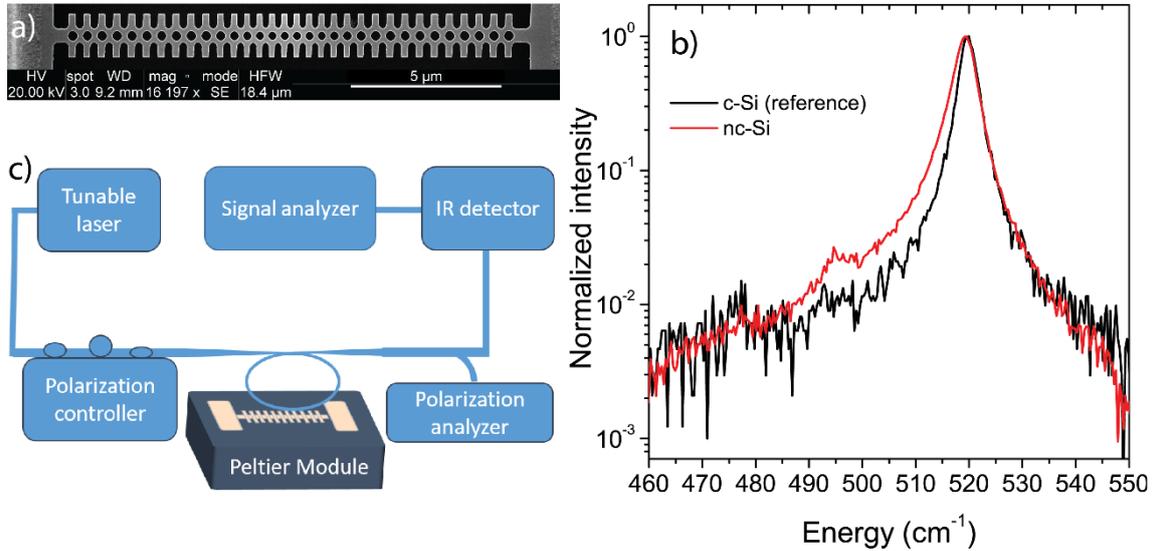

*Figure 1. a) SEM image of an OMC in nc-Si with ratio between the geometrical parameters of the mirror and central cells of 0.8. b) Raman spectra of a monocrystalline silicon sample (black) and the nc-Si layer (red) c) Scheme of the experimental setup used for characterizing the OMC.*

The following experiments were performed in a standard set-up to characterize the optical and mechanical properties of nc-Si OMCs [18] (see (Fig. 1(c)). A tuneable infrared laser, the wavelength of which ($\lambda_{laser}$) covers the range between 1.44–1.64 µm, is connected to a tapered microlooped fibre. The polarization of the light entering the tapered region is set with a polarization controller. The thinnest part of the tapered fibre is placed parallel to the OMC, in contact with an edge of the etched frame. The gap between the fibre and the structure is about 0.2 µm. A polarization analyser is placed after the tapered fibre region. The long tail of the evanescent field locally excites the resonant optical modes of the OMCs. Once in resonance, the mechanical motion activated by the thermal Langevin force causes the transmitted intensity to be modulated around the static value. To check for the presence of a radio-frequency modulation of the transmitted signal an InGaAs fast photoreceiver with a bandwidth of 12 GHz was used. The radio-frequency voltage is connected to the 50 Ω input impedance of a signal analyser with a bandwidth of 13.5 GHz. All the measurements were performed in an antivibration cage at atmospheric conditions of air pressure and temperature.

In Fig. 2(a) we show a typical example of an optical transmission spectrum of one of the fabricated nc-Si OMCs. Several sharp optical resonances corresponding to TE-like confined modes appear superposed to an oscillating transmission associated to whispering gallery modes of the microlooped fibre. In the best cases, the resonances display overall optical quality factors of about $Q_o=1.16\times10^4$ (inset of Fig. 2(a)), i.e., an overall decay rate of $\kappa=1.2\times10^{12}$ s$^{-1}$. The intrinsic optical quality factor ($Q_{o,i}$) as calculated from $Q_o$ and the transmitted fraction[2] is then $Q_{o,i}=1.55\times10^4$, meaning an intrinsic and extrinsic optical decay rates of $\kappa_i=8\times10^{11}$ s$^{-1}$ and $\kappa_e=3\times10^{11}$ s$^{-1}$ respectively. It is worth to note that identical designs fabricated in c-Si give optical $Q_o$ values that are about 5 times larger,[31] meaning that material losses are a dominant mechanism in nc-Si OMCs.

In Fig. 2(b) we compare the thermo-optic (TO) contribution of two nominally-equivalent OMCs fabricated on c-Si (top panel) and on nc-Si (bottom panel). It is worth noting that, in this specific analysis, we neglect other optical nonlinearities such as free-carrier-dispersion or Kerr effects because of their negligible contribution in comparison to that of TO. The black curves show the spectral shape of the first resonance observed in each OMC, thus corresponding to first order modes. The same measurement at high input power shows asymmetric spectra caused by the red-shift of the resonance position ($\lambda_r$) due to TO dispersion[32] in response to an effective temperature increase ($\Delta T$) of the OMC. In this regime, decreasing the laser-cavity detuning from the blue-side results in an increase of the intracavity photon number ($n_o$). The maximum value ($n_{o,max}$) is achieved when the laser-cavity detuning vanishes, i.e., at the transmission minimum. Given that, for a two-sided cavity, $n_{o,max}$ can be expressed as $n_{o,max}=2P_{in}(\kappa_e/(\kappa^2)\lambda_r/hc$, where $P_{in}$ is the input laser power, $P_{in}$ has been adjusted to obtain the same value of $n_{o,max}$ for both OMCs. This procedure allows a fair comparison between the TO contribution of each OMC. Red curves of Fig. 2(b) show that, for $n_{o,max}=2.3\times10^4$, we obtain $(\lambda_r(n_{o,max})-\lambda_{r,o})_{c-Si}=1.0$ nm and $(\lambda_r(n_{o,max})-\lambda_o)_{nc-Si}=11.3$ nm for c-Si and nc-Si respectively. The TO coefficients ($\partial\lambda_{r,o}/\partial\Delta T$) were determined in an independent way by quantifying the resonance optical shift at low laser power as a function of $\Delta T$, which was set by a Peltier placed below the sample. Interestingly, we have obtained the same $\partial\lambda_{r,o}/\partial\Delta T$ values for equivalent OMC geometries fabricated in c-Si and in nc-Si (see Figure S1). With this procedure, we have obtained that the optical shift dependence is linear with $\Delta T$ up to the maximum value achieved by the Peltier ($\Delta T=70K$) with a slope given by $\partial\lambda_r/\partial\Delta T=0.09nm/K$. Therefore, the difference observed in $\lambda_r(n_{o,max})-\lambda_{r,o}$ between both material platforms is directly associated to a much higher $\Delta T$ in the case of nc-Si ($\Delta T(n_{o,max})=126K$) than in the case of c-Si ($\Delta T(n_{o,max})=11K$). As it will be discussed further below, this is a consequence of both a lower heat dissipation rate of the nc-Si material and of a faster heating rate associated to an efficient Two-Step-Absorption (TSA) mechanism involving mid-gap states.

When the excitation laser is on resonance, the thermally activated mechanical modes displaying significant OM coupling induce fluctuations on the optical resonance position. As a consequence, a rich RF spectrum is observed (Fig. 2(c)), the various peaks corresponding to mechanical modes distributed over a spectral range between few MHz and several GHz. The RF signal observed in the region below 1 GHz is mostly associated to the extended mechanical modes involving the oscillation of the whole beam. Although the expected single-particle OM coupling rate ($g_o$) of this kind of modes is expected to be low, small fabrication asymmetries disturb the spatial distribution of the optical mode, leading to significant enhancement of the experimental $g_o$.[4] The intrinsic mechanical quality factor ($Q_{m,i}$) spectrum of these modes is on the order of $10^2$, being probably limited by thermo-elastic and/or visco-elastic loss mechanisms. On the other hand, rather large values of $Q_{m,i}$ are found for the localized modes present above 1 GHz. In particular, the mechanical mode appearing at 2.63 GHz (black curve of the inset of Fig. 2(c)) displays a maximum $Q_{m,i}$=1.95x10$^3$. This value is higher than what found in equivalent c-Si OMCs at room temperature,[10,18] which is $Q_{m,i}$= 0.67x10$^3$ (green curve of the inset of Fig. 2(c)). The dominant damping mechanism for the modes above 1 GHz is probably thermo-elastic and scattering with thermal phonons,[33] grains and boundaries.[34] It is worth noting that $g_o$ values are compatible with what has been reported previously for c-Si,[31] i.e., on the order of the few hundred kHz for the most intense modes. It is then reasonable to state that the photoelastic coefficients of the nc-Si material presented here are similar to those of c-Si.

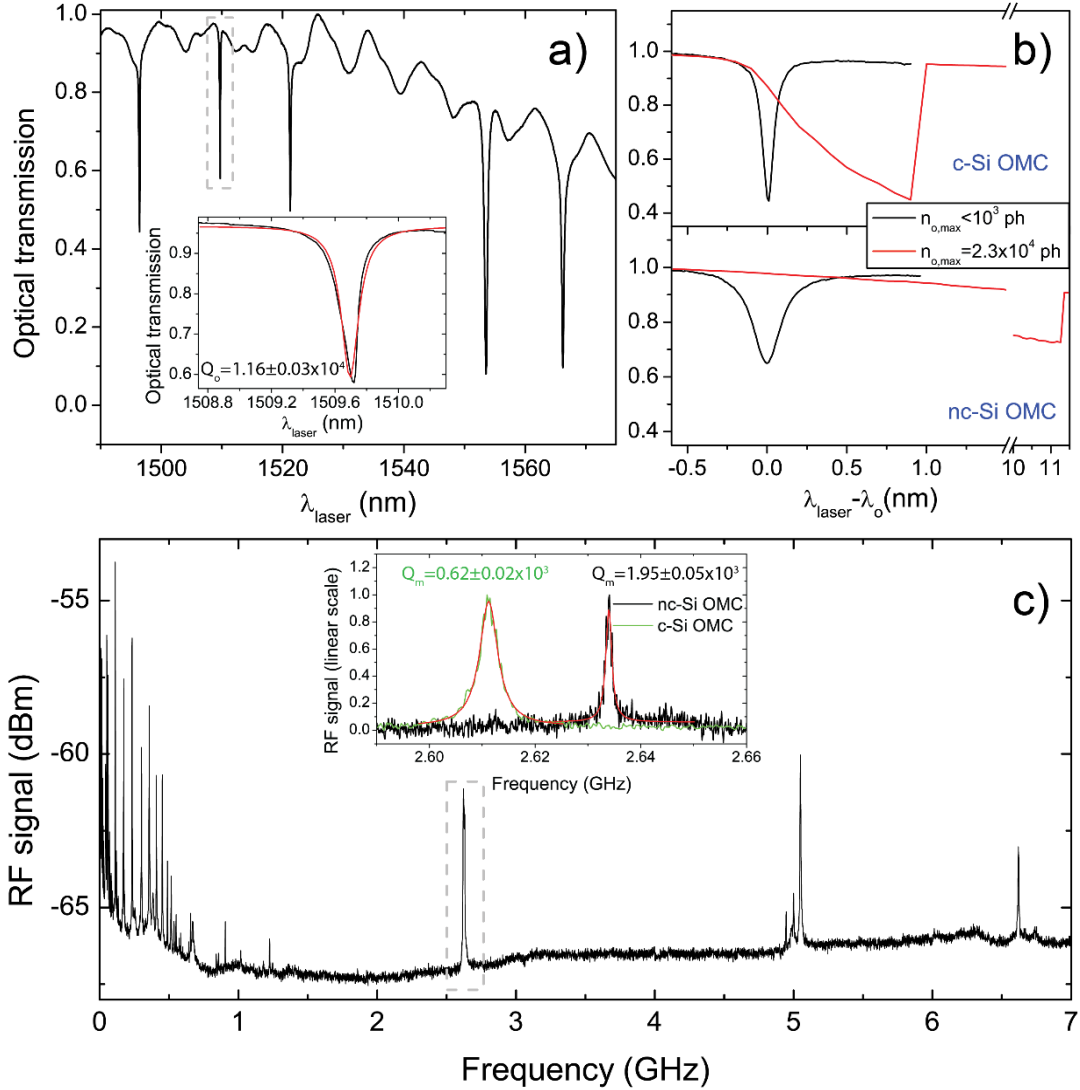

***Figure 2. a)*** *Optical transmission spectrum of one of the OMCs fabricated of nc-Si. On the inset we show a zoom on one of the optical resonances together with a lorentzian fit of the resonance (in red).* ***b)*** *Optical spectrum of the first order optical resonance of a c-Si (top panel) and a nc-Si (bottom panel) OMCs for two different values of the intracavity photon number. The x-axis is referred as the detuning between $\lambda_{laser}$ and the unperturbed resonance position ($\lambda_{r,o}$).* ***c)*** *RF spectrum obtained by exciting the cavity with the optical mode highlighted in panel a). On the inset we show a zoom a mechanical mode appearing at 2.63 GHz (black) and a mode associated to an equivalent c-Si OMCs (green) together with lorentzian fits of the resonances (in red).*

The OMCs fabricated on nc-Si do not currently operate in the resolved-sideband regime ($\Omega_m/\kappa_i \approx 0.5 < 1$, where $\Omega_m$ is the mechanical frequency), in which radiation-pressure is efficient enough to generate significant dynamical back-action effects. However, large energies stored in these OMCs give rise to other strong optical non-linearities involving the OM interaction. In particular, in this work we address the so-called self-pulsing (SP), which arises from the dynamical interplay between free-carrier-dispersion and the TO effect. The SP mechanism has been reported before by our and other groups in c-Si based OMCs and a detailed description of the phenomenon can be found elsewhere.[3] As a result of the SP, the cavity resonance oscillates periodically around the laser line at a frequency denoted by $v_{SP}$, thus impinging an anharmonic modulation of the radiation pressure force within the cavity. If one of the frequency harmonics of the force is resonant with a mechanical mode displaying a significant value of $g_o$, that particular mode can be driven into the so-called phonon lasing regime and trigger complex non-linear dynamics such as chaos.[4] Figure 3 shows a 2D plot of the RF spectra obtained as a function of the laser wavelength for a given input laser power of 4 mW. The threshold to trigger the SP dynamics is found slightly beyond the vertical grey line, below which it is possible to transduce the thermally activated mechanical motion of the extended modes (right panel, grey curve), similarly to what reported in Fig. 2(c). By increasing the laser wavelength from the threshold, $v_{SP}$ increases accordingly and the OMC can enter into a phonon lasing regime for different mechanical modes depending on which one is resonant with a harmonic of $v_{SP}$. This appears in Fig. 3 as frequency plateaus, where the linewidth of the RF peaks decrease dramatically in response to the robust bidirectional coupling between the lasing modes and the SP by means of the OM interaction.[3] Maximum values of $v_{SP}$ reach about 0.3 GHz, which is more than 5 times faster than reported for c-Si OMCs (0.054 MHz in Ref. 3). Under the experimental conditions to activate such a fast SP, it is possible to drive resonantly the mechanical mode at 0.302 GHz into the phonon lasing regime (vertical green line and green curve of Fig. 3). Interestingly, the SP dynamics remains active while scanning the excitation laser wavelength for about 30 nm, which is several times larger than what can be achieved in c-Si.[3] It is worth noting that similar laser powers are used for both material platforms.

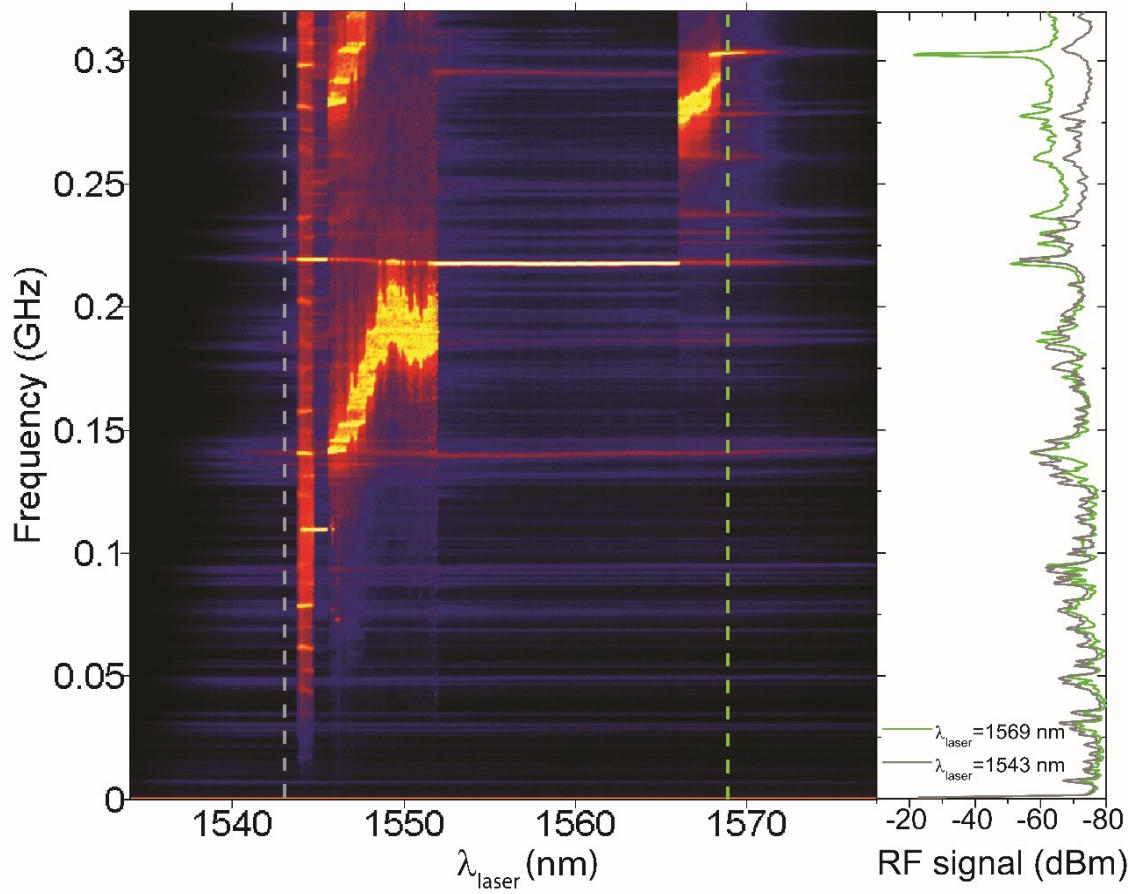

*Figure 3. Non-linear dynamics of the nc-Si OMC. The left panel shows the colour contour plot of the RF spectrum as a function of the laser wavelength obtained at about 4 mW. The right panel shows the specific RF spectra at the two laser wavelength highlighted by vertical dashed lines in the main panel. The grey curve corresponds to the transduction of thermally activated mechanical modes. The green curve reports a mechanical mode placed at 0.302 GHz that is driven into the phonon lasing regime.*

In order to interpret the experimental results and understand the differences with respect to the c-Si case we have analysed the behaviour of the isolated SP (uncoupled to the mechanics) by numerically solving the system of two coupled first order differential equations describing the dynamics of the free carrier population ($N$) and $\Delta T$. [3]

$$\dot{N} = \quad + \beta n_o^2 \quad (1a),$$

$$\dot{\Delta T} = -\Gamma_{th}\Delta T + \alpha_{fc} N n_o \quad (1b)$$

where $\alpha_{fc}$ is defined as the rate of temperature increase per photon and unit free-carrier density and $\beta$ is a coefficient accounting for Two-Photon-Absorption (TPA) and TSA. Concerning TSA, this mechanism is generally neglected in c-Si, since there is a rather low density of mid-gap localized defect states. On the contrary, the density of these states is much higher in nc-Si, thus playing an important role enhancing light absorption below the gap energy and the free-carrier recombination rates ($\Gamma_{fc}$). [24]

The degree of tuneability of $\nu_{SP}$ depends on the excitation laser parameters (wavelength and power), on $Q_o$, on the rate at which heat is dissipated out of the cavity ($\Gamma_{th}$) and on $\Gamma_{fc}$. In Figure 4(a) we report the maximum $\nu_{SP}$ ($\nu_{SP,max}$) for different values of $\Gamma_{fc}$ and $Q_o$. As a starting point we have taken the set of parameters used in Ref. 3 leading to a good agreement with the dynamics observed in c-Si based OMCs and a $\nu_{SP,max}$=0.054 GHz (first black dot in Figure 4a). Thus, the starting parameters are $\Gamma_{fc}$=2 GHz (extracted from Ref. 22), $Q_o$=2.2x10$^4$ (extracted from our experimental results) and $\Gamma_{th}$=2 MHz (adjusted to fit the experimental SP dynamics). The strongest dependence of $\nu_{SP,max}$ has been found by varying $\Gamma_{fc}$, where $\nu_{SP,max}$ increases linearly. The behaviour with $Q_o$ is less evident as at low $\Gamma_{fc}$ values $\nu_{SP,max}$ decreases with $Q_o$, while at high $\Gamma_{fc}$ values there is no clear correlation. In the latter case, if $Q_o$ is low enough, the coupled system does not display a Hopf bifurcation and the solution is a stable fixed point. It is important to note that blue dots correspond to solutions obtained by using the experimental value of $Q_o$ for nc-Si.

Finally, we have also studied the influence of varying $\Gamma_{th}$ on the SP dynamics. In this regard we have concluded that $\nu_{SP,max}$ is independent from $\Gamma_{th}$. To illustrate this, Figures 4b and 4c show the Fourier transform of the simulated optical transmission as a function of the laser wavelength for two different values of $\Gamma_{th}$ ($\Gamma_{th}$=2 MHz and $\Gamma_{th}$=1 MHz respectively), $\Gamma_{fc}$ and $Q_o$ being those of the black dot highlighted in Fig. 4a. In both simulated cases $\nu_{SP,max}$ =0.2 GHz, although the SP threshold is achieved at shorter laser wavelengths for higher $\Gamma_{th}$ values. It is also clear from our simulations that increasing $\Gamma_{th}$ leads to a decreasing of the laser bandwidth in which SP is active. Experimentally, in nc-Si we observe both high $\nu_{SP,max}$ values and wider laser bandwidths in which SP is active. Thus, the differences observed experimentally between the two material platforms are associated to a lower $\Gamma_{th}$ for nc-Si, which is an acknowledged fact [30] in agreement with the results obtained when quantifying the TO effect, and a higher $\Gamma_{fc}$ ( about $\Gamma_{fc}$ =7 GHz). The latter value of $\Gamma_{fc}$ is consistent with that reported by Preston et al. [24] ($\Gamma_{fc}$ =7.4 GHz) when demonstrating high-speed polycrystalline silicon based light modulators. As stated before, the enhancement of $\Gamma_{fc}$ in nc-Si is associated to a high-density of mid-gap localized defect states, which act as recombination centers for the free-carriers. [24]

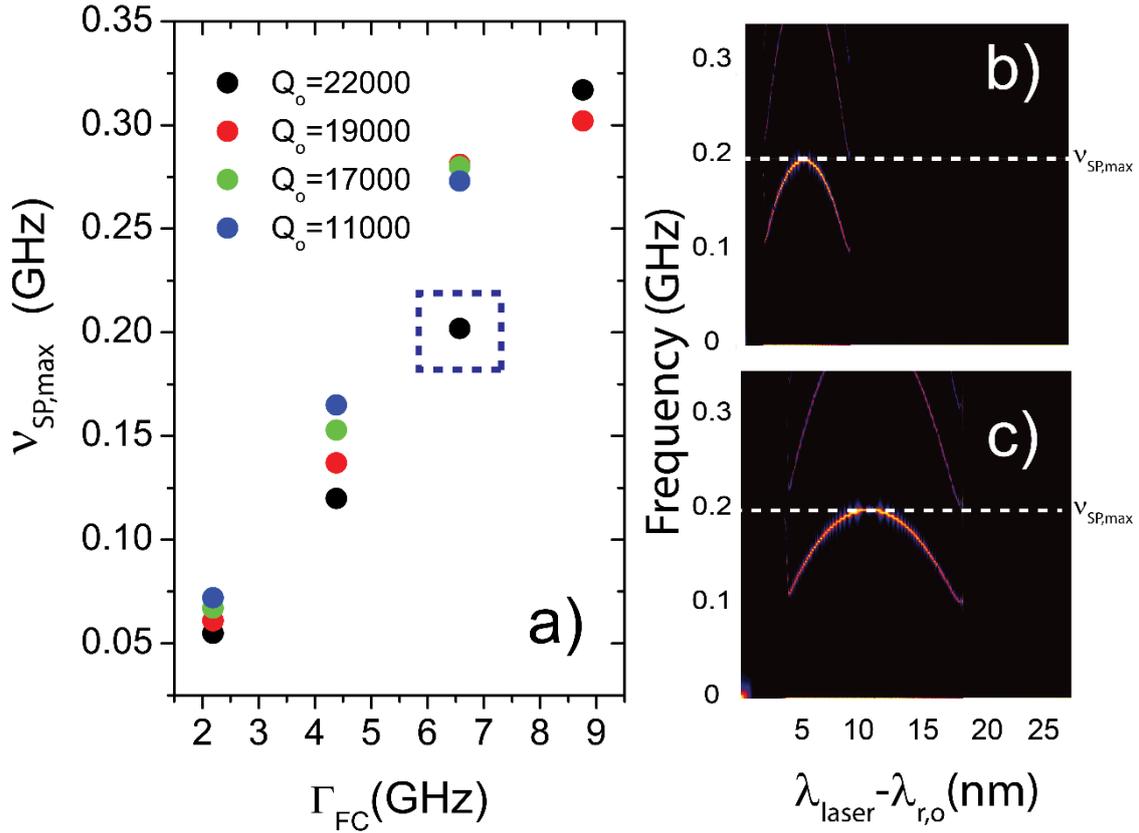

***Figure 4.*** *Numerical simulations of the self-pulsing (SP) dynamics in the nc-Si OMC. **a)** Maximum SP frequency as a function of the free-carrier relaxation rate. The different colours are associated to different values of the quality factor of the optical cavity. Panels **b)** and **c)** show the simulated radio-frequency spectrum of the transmitted signal as a function of the laser wavelength for two different values of the rate at which heat is dissipated out of the cavity ($\Gamma_{th}$=2MHz and $\Gamma_{th}$=1MHz for panels b) and c) respectively).*

At higher laser powers of ~8 mW, it is also possible to reveal another bifurcation in which the dynamical system switches abruptly from a phonon lasing state (red curve in Fig. 5) to a chaotic regime (green curve in Fig. 5), similarly to what was reported in Ref. 4 for c-Si OMCs. In the chaotic regime the RF spectrum is broad with a rich structure of sharp RF peaks associated to the activation of different mechanical modes (see the correspondence with the spectrum of the thermally activated mechanical peaks of the black curve in Fig. 5) due to the broadband nature of the optical force. Numerical simulations of the system reveal that chaos is only present in the SP equation, while the harmonic oscillators stay coherent. [4]

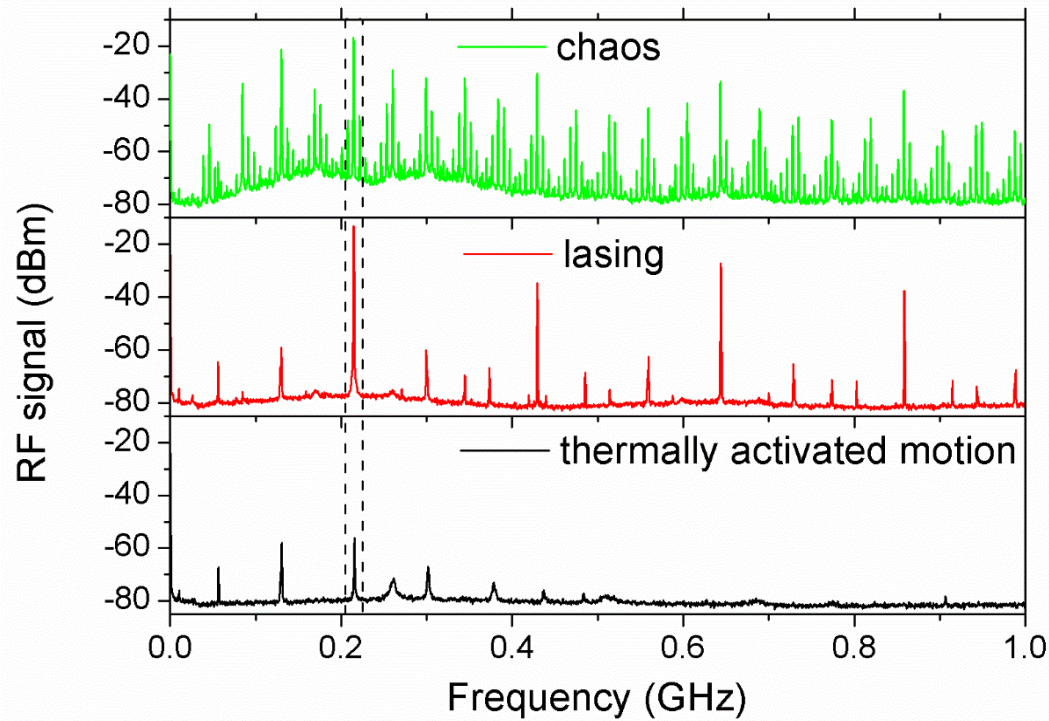

*Figure 5.* RF spectrum of thermally activated motion (black), phonon lasing at 0.21 GHz (red) and chaotic behaviour (green). The three different spectra have been obtained by exciting the same optical mode of a nc-Si OMC using different laser wavelengths at 8 mW laser power.

Conclusions

In this work we have demonstrated experimentally (– for the first time to our knowledge –) the coupling of GHz mechanical modes into optical waves as well as non-linear phenomena such as self-pulsing, phonon lasing and chaos in an nc-Si OMC. Although the optical Q factor is somewhat reduced in comparison to an identical c-Si cavity, it is large enough as to enable the transduction of GHz vibrations into the output optical signal and the activation of self-pulsing dynamics. Strong thermal effects in nc-Si allow us to observe an extremely large (30 nm) tuning range of the optical resonances. In addition, we have reported a factor of five enhancement of the maximum frequencies of the dynamic solutions of the nc-Si based OMC, which is mostly associated to a higher free-carrier relaxation rate due to the presence of mid-gap localized defect states.

The use of nc-Si has some practical advantages over c-Si for optical and optomechanical applications. For instance, nc-Si allows for the development of multilayer OMCs, in which a strong and tuneable optomechanical interaction is expected,[35] by adding extra degrees of freedom for coupling multiple cavities. In addition, the thickness of the silicon layer can be adjusted freely, so thicker silicon slabs required for full phononic bandgaps in 2D honeycomb lattices can be realised.[36,37] Finally, nc-Si wafers are less expensive than SOI wafers and, consequently, their use would be beneficial for industrial exploitation of OMC chips.

Acknowledgements

This work was supported by the European Commission project PHENOMEN (H2020-EU-713450), the Spanish Severo Ochoa Excellence program and the MINECO project PHENTOM (FIS2015-70862-P). ICN2 authors acknowledge support from the Severo Ochoa Program (MINECO, Grant SEV-2013-0295) and funding from the CERCA Programme / Generalitat de Catalunya. DNU gratefully acknowledges the support of a Ramón y Cajal postdoctoral fellowship (RYC-2014-15392). We thank M. Blomberg for fruitful discussions and L. Nurminen for the fabrication of the nc-Si wafers.

**Supplementary Material**

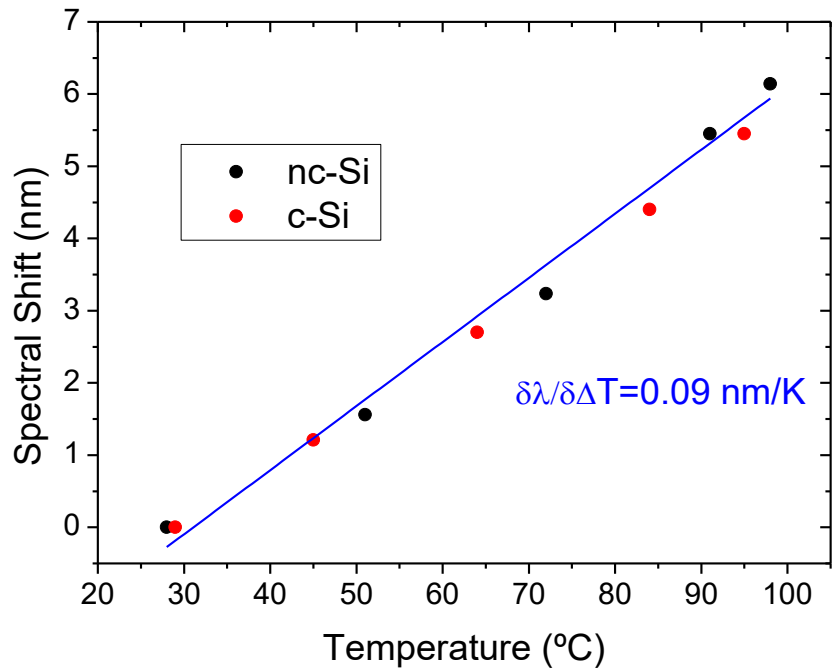

*Figure S1. Calibration of the TO coefficient. Spectral shift of the first optical resonance of an OMC as a function of the Peltier temperature placed below the sample. Black and red dots correspond to a nc-Si and a c-Si material respectively. The OMCs are geometrically equivalent, the defect region being constructed with Λ=0.8.*